\documentclass{article}
\usepackage{spconf,amsmath,graphicx}
\usepackage{multirow}
\usepackage{booktabs}

\title{Rep Works in Speaker Verification}
\name{Yufeng Ma$^1$, Miao Zhao$^1$, Yiwei Ding$^{1,2}$, Yu Zheng$^1$, Min Liu$^1$, Minqiang Xu$^1$\textsuperscript{*}}
\address{$^1$ SpeakIn Technologies Co. Ltd. \\ $^2$ Fudan University}
\begin{document}
\ninept
\maketitle

\begingroup\renewcommand\thefootnote{*}
\footnotetext{Corresponding author}
\begin{abstract}
Multi-branch convolutional neural network architecture has raised lots of attention in speaker verification since the aggregation of multiple parallel branches can significantly improve performance. However, this design is not efficient enough during the inference time due to the increase of model parameters and extra operations. In this paper, we present a new multi-branch network architecture RepSPKNet that uses a re-parameterization technique. With this technique, our backbone model contains an efficient VGG-like inference state while its training state is a complicated multi-branch structure. We first introduce the specific structure of RepVGG into speaker verification and propose several variants of this structure. The performance is evaluated on VoxCeleb-based test sets. We demonstrate that both the branch diversity and the branch capacity play important roles in RepSPKNet designing. Our RepSPKNet achieves state-of-the-art performance with a 1.5982\% EER and a 0.1374 minDCF on VoxCeleb1-H.
\end{abstract}
\begin{keywords}
speaker verification, speaker recognition, re-parameterization
\end{keywords}
\section{Introduction}
\label{sec:intro}

Speaker verification aims to verify a speaker's identity given an audio segment. In recent years, deep neural networks (DNNs) has improved the performance of speaker verification systems which outperform the traditional i-vector system \cite{ivector}. Most DNN-based systems, such as x-vector \cite{snyder2018x}, r-vector \cite{zeinali2019but}, and the recently proposed ECAPA-TDNN \cite{ecapa_first, thienpondt2020idlab},  consist of three parts:  (1) a network backbone to extract frame-level speaker representations, (2) a pooling layer to aggregate the frame-level information, and (3) a loss function. This paper focuses on the backbone architecture, which is the core part of the DNN models.

The backbone architecture can be a 1-dimensional convolutional neural network (TDNN) \cite{snyder2018x}, a 2-dimensional convolutional neural network (CNN) \cite{zeinali2019but, zhou2020resnext}, a recurrent neural network (RNN), and even a hybrid architecture that combines TDNN, CNN, RNN, and Transformer-like structures \cite{zhu21c_interspeech}. Several modifications of the backbone architecture are made to improve the performance. These modifications include adding a channel attention \cite{Hu_2018_CVPR}, transforming the custom convolution into a multi-scale convolution \cite{Xie_2017_CVPR, Gao_2021}, and aggregating multi-layer or multi-stage features \cite{ecapa_first, Jung_2020msea}. However, all these methods above only focus on the improvement of single-branch structures and neglect a multi-branch way of designing neural networks. Adding parallel branches \cite{szegedy2014going, szegedy2015rethink, Szegedy2017Inceptionv4IA} can significantly enlarge the model capacity and enrich the feature space, which results in better model performance. Yu \textit{et al.} \cite{YuL20dtdnn} proposed a multi branch version of densely connected TDNN structure with a selective kernel (D-TDNN-SS) and this model achieved competitive performance in speaker verification.

Though the complicated multi-branch structure has proved its power, more parameters and connections usually lead to a slow inference speed. Recent researches \cite{Ding_2019_ICCV, ding2021repvgg, ding2021diverse} proposed a new technique called re-parameterization to solve the increasing inference cost. The main idea of this technique is to design a training time multi-branch structure which can be transformed to a single path with only one custom convolution during the inference time. This technique decouples training time and inference time architecture and ensures that the output remains the same. Inspired by this re-parameterization, we \cite{zhao2021speakin} first introduced the original RepVGG model into the speaker verification and obtained first place in both Track 1 and Track 2 of VoxSRC2021.

The spectrogram whose shape is $\ C \times F \times T$ is different from the image data that commonly serve as data input in computer vision tasks. $C$ here means the feature maps (channel), $F$ means the frequency features and $T$ means the time axis. We presented that the design of the multi-branch structure can be heuristic due to this difference of input data. The structure is task-specific and various re-parameterizable branches achieve different performances. To fully investigate how this re-parameterization works in speaker verification, we followed the work in \cite{ding2021repvgg, zhao2021speakin} and proposed several variants of the original RepVGG block. We evaluated the performance of these systems on Voxceleb1-O, VoxCeleb1-E, and Voxceleb1-H \cite{nagrani2017voxceleb, chung2018voxceleb2}. Based on these results, we at the first time proposed a new re-parameterizable structure named RepSPKNet and demonstrated the importance of branch diversity and branch capacity in designing multi-branch structures. Our RepSPKNet can be transformed to a stack of simple $K \times K$ convolutions and ReLU layers during the inference time which results in a fast inference speed and competitive performance. The proposed RepSPKNet model achieved a 1.5982\% EER and a 0.1374 minDCF on VoxCeleb1-H.

The paper is organized as follows: Section \ref{sec:rep} reviews the prior works related to re-parameterization. Section \ref{sec:model_archi} presents our baseline system with a RepVGG backbone and other variants. In section \ref{sec:exp_result}, we discuss the experiment details and analyze the result. The analysis finally derives our carefully designed RepSPKNet. Section \ref{sec:conclusion} concludes this paper.

\section{Re-parameterization}
\label{sec:rep}

Structural re-parameterization is used to avoid the extra parallel branch parameters and the slow inference speed via converting a multi-branch structure into a single path. ACNet \cite{Ding_2019_ICCV} proposed a $1\times3$ kernel convolution with batch normalization ($1\times3$ CONV-BN) and a $3\times1$ CONV-BN to strengthen the original $3\times3$ CONV-BN. RepVGG added a $1\times1$ CONV-BN and an identity batch normalization layer (ID-BN) in parallel. Furthermore, DBB \cite{ding2021diverse} proposed a diverse branch block and gave more general transformations. It is worth mentioning that the combinations of these transformations also satisfy the request for re-parameterization. Here we list the general transformations as follows:
\begin{figure}[t]
  \centering
  \includegraphics[width=0.21\textwidth,height=0.50\textwidth]{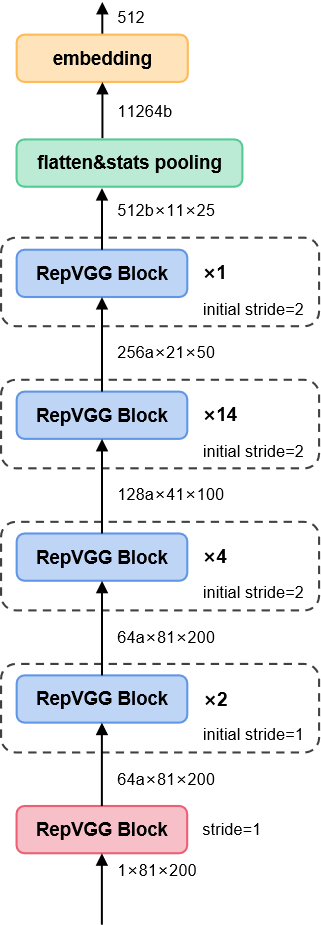}
  \caption{The baseline system. The $a$, $b$ here denote the layer width parameter. Initial stride means the stride of first block of each stage. The input format is $C\times F\times T$.}
  \label{fig:rep_baseline}
\end{figure}
\begin{itemize}
    \item \textbf{CONV-BN fusion} Batch normalization can be fused into its preceding convolution. Given a kernel weight $F$, the fused parameters can be formulated as:
    \begin{equation}
        F_i^{'} = \frac{\gamma_i}{\sigma_i}F_i,\  b_i^{'} = -\frac{\mu_i\gamma_i}{\sigma_i}+\beta_i
    \end{equation}
    where \textit{i} denotes the \textit{i}-th channel, and $\gamma$, $\mu$, $\sigma$, $\beta$ denote the scaling factor, mean, variance and bias of the BN layer.
    \item \textbf{Parallel conv addition} Convolutions with different kernel size in different branches can be fused into one convolution by zero-padding small kernels and applying a simple element-wise addition.
    \item \textbf{Sequential convolutions fusion} A sequence of $1\times1$ CONV-BN and $k\times k$ CONV-BN can be fused into a $k\times k$ CONV whose parameters can be formulated as:
    \begin{equation}
        F^{'} = F^{(2)} \ast TRANS(F^{(1)})
    \end{equation}
    \begin{equation}
        b_i^{'}=\sum\limits_{d=1}^{D}\sum\limits_{u=1}^{K}\sum\limits_{v=1}^{K}b_d^{(1)}F_{i,d,u,v}^{(2)}
    \end{equation}
    where $F^{(1)}$, $b^{(1)}$ and $F^{(2)}$, $b^{(2)}$ denote the weights and bias of convolutions, and $TRANS$ denotes transpose operation.
    \item \textbf{Average pooling transformation} A kernel $k$ average pooling can be transformed to a $k\times k$ convolution. The parameter is
    \begin{equation}
        F^{'} = \frac{1}{k^2}I_{k\times k}
    \end{equation}
    where $I_{k\times k}$ is an identity matrix.
\end{itemize}

\section{Our Proposed RepSPKNet system}
\label{sec:model_archi}
The RepVGG based speaker verification system has already shown its competitive performance \cite{zhao2021speakin}. This section describes our baseline system and our proposed variants.
\subsection{Baseline system}
\begin{figure}[t]
  \centering
  \includegraphics[width=0.48\textwidth,height=0.19\textwidth]{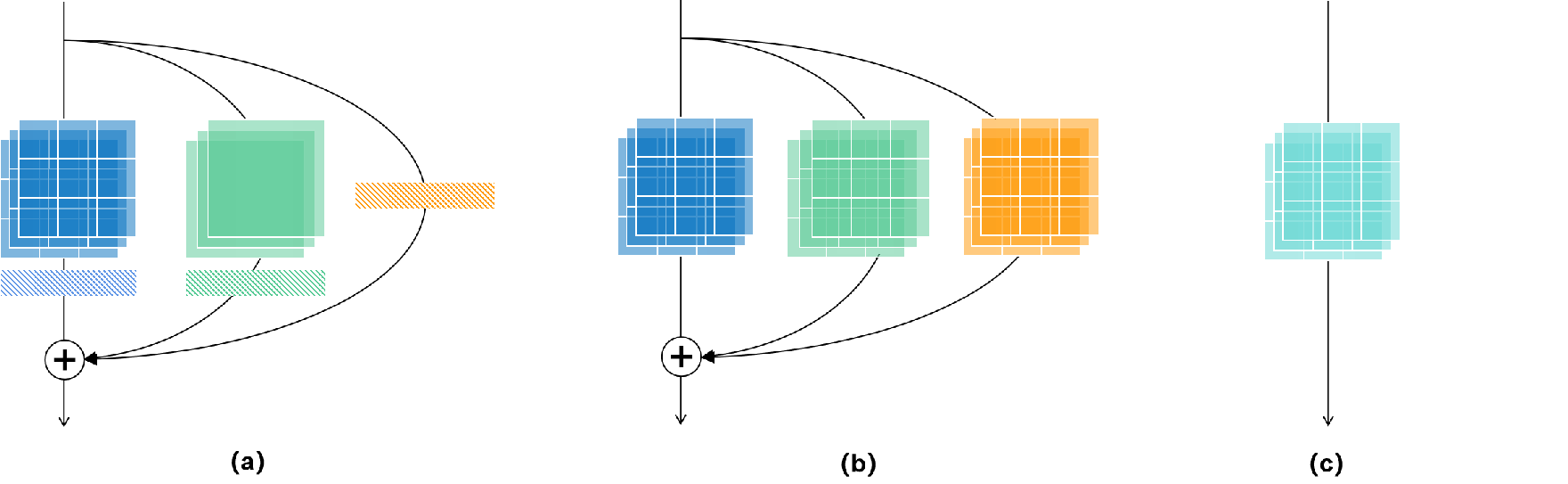}
  \caption{Architecture of RepVGG block. Here (a) is the training time state. (b) demonstrates the process of CONV-BN fusion. (c) is the inference time state. $\oplus$ denotes element-wise addition. A ReLU is added after branch addition.}
  \label{repvgg_block}
\end{figure}
Here we present our baseline system that consists of a RepVGG-A backbone, a statistical pooling \cite{snyder2017deep}, a 512-dimensional embedding layer and an additive margin softmax (AM-Softmax) loss function \cite{wang2018cosface, wang2018additive}. The detailed topology is shown in \textbf{Fig. \ref{fig:rep_baseline}}. 

As \textbf{Fig. \ref{repvgg_block}} shows, the basic RepVGG block consists of three parallel branches: (1) a $3\times3$ CONV-BN, (2) a $1\times1$ CONV-BN, and (3) an ID-BN. The ID-BN branch exists only when the input channel equals the output channel. According to the transformations in Section \ref{sec:rep}, it is easy to verify that these three branches can be merged into one $3\times3$ convolution during the inference time.  The RepVGG-A backbone consists of a stem layer and four stages. These stages contain 2, 4, 14, and 1 RepVGG blocks respectively and the stem layer is also a RepVGG block. The complexity of our backbone depends on the layer width parameters (\textit{a}, \textit{b}). For RepVGG-A0, we set $a=0.75$, and $b=2.5$. For RepVGG-A1, $a = 1.0$, and $b=2.5$. For RepVGG-A2, $a=1.5$, and $b=2.75$. We slightly change the original stride setting \cite{ding2021repvgg} to make this backbone fit the speaker verification task. Both the first stage and the stem layer have a stride of 1. The other stages have a stride of 2. The format of input is $1 \times F \times T$. The output of the backbone has a shape as $512b \times \frac{F}{8} \times \frac{T}{8}$ and is reshaped to $(512b \times \frac{F}{8}) \times \frac{T}{8}$. The whole backbone can be transformed to a stack of $3\times3$ convolutions and ReLU layers during the inference time.

A statistical pooling layer is applied to aggregate the speaker information. We calculate the mean and the standard deviation of the backbone output along the time axis. The mean and standard deviation are concatenated and then compressed to a 512-dimensional vector which serves as the speaker embedding. The AM-Sofmtax is used to classify speakers which can be formulated as:
\begin{equation}
\mathcal{L}_{AM'}\!=\!- \frac{1}{N}\!\sum_{i=1}^{N}log \frac{e^{s\cdot (cos\theta_{i,y_i}-m)}}
  {e^{s\cdot (cos\theta_{i,y_i}-m)}+\!\!\!\sum\limits_{j=1,j \neq y_i}^{C}e^{s\cdot cos\theta_{i,j}}}
  \label{eq4}
\end{equation}
where N denotes the number of samples, C denotes the number of speakers, s denotes the scaling factor, m denotes the margin penalty and $cos\theta_{i,j}$ denotes the angle between weight vector $w_j$ and the \textit{i}-th sample.

\subsection{Variants of the baseline system}
\label{variants}
\begin{figure}[h]
  \centering
  \includegraphics[width=0.48\textwidth,height=0.17\textwidth]{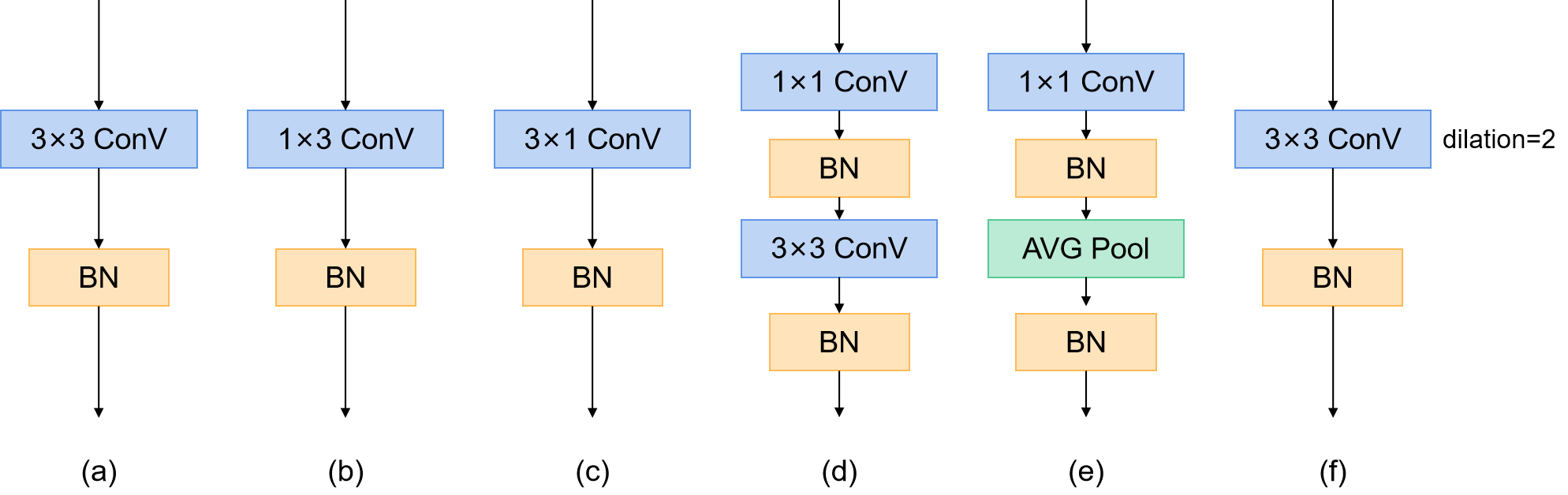}
  \caption{Variants of the original RepVGG basic block. For convenience, the $3\times 3$ CONV-BN and the ID-BN not depicted.}
  \label{fig:variants}
\end{figure}
As we introduced in Section \ref{sec:intro}, the original architecture of this RepVGG block is designed for computer vision tasks. Though this architecture has proved its superiority, it is task-specific and may not be the optimal structure in speaker verification. The $3\times3$ CONV-BN is the main branch of this architecture, while the ID-BN serves as a residual connection to avoid the gradient vanishing problem. To investigate the re-parameterization in speaker verification, we fixed the $3\times3$ CONV-BN and ID-BN branches and proposed several variants to replace the $1\times 1$ CONV-BN. As \textbf{Fig. \ref{fig:variants}} demonstrates, structure (a) is a duplicate $3\times3$ CONV-BN. Structure (b) and (c) are borrowed from ACNet. Structure (d) and (e) are borrowed from DBB. Structure (f) consists of a $3\times3$ convolution with a dilation of 2 and a batch normalization layer. The original RepVGG blocks are replaced with these variants of the original block to form new speaker verification models. According to the transformations in Section \ref{sec:rep}, all these variants (except \textbf{(f)}) can be transformed to a $3\times3$ convolution which means the inference bodies of these new models remain the same when compared to the original RepVGG inference time state.

\section{Experiments and results}
\label{sec:exp_result}
\subsection{Dataset and features}
All our models adopted the VoxCeleb2 development set \cite{chung2018voxceleb2} as our training set. This dataset contains 1,092,009 utterances and 5,994 speakers in total. Our data augmentation consisted of two parts: (1) A \textbf{3-fold speed augmentation} \cite{zhao2021speakin, yamamoto2019speaker} was implemented at first to generate extra twice speakers based on the SoX speed function. (2) We followed the data augmentation method provided by the Kaldi VoxCeleb recipe. The RIRs \cite{ko2017study} and MUSAN \cite{snyder2015musan} dataset was used.

After the augmentation process, 16,380,135 utterances from 17,982 speakers were generated. We extracted 81-dimensional log Mel filter bank energies based on Kaldi without voice activity detection (VAD). The window size is 25 ms, and the frame-shift is 10 ms. All the features were cepstral mean normalized.

\subsection{Experiment setup}
200 frames of each sample in one batch were randomly selected. The SGD optimizer with a momentum of 0.9 and a weight decay of \textit{1e-3} was used. We used 8 GPUs with mini-batch as 1,024 and an initial learning rate of 0.08. We adopted ReduceLROnPlateau scheduler and the minimum learning rate is \textit{1e-6}. The margin of the AM-Softmax loss is set to \textit{0.2} and the scale is 36. All our systems were evaluated on VoxCeleb1-O, VoxCeleb1-E, and VoxCeleb1-H. Trials were scored by cosine similarity of the 512-dimensional embeddings and no score normalization was implemented. The criterion is equal error rate (EER) and minimum decision cost function (DCF) where $C_{FA} = 1$, $C_{M} = 1$, and $p_{target} = 0.01$.

\subsection{Results and analysis}
\subsubsection{Ablation study of base model}
\begin{table}[t]
  \centering
  \caption{Ablation study on our baseline model RepVGG-A0. For convenience, we omitted the \% sign of EER and used Var to denote the variant using the structures we proposed. Var \textit{f} cannot be transformed to a custom $3\times3$ convolution.}
  \label{tab:tabel_one}
  \setlength{\tabcolsep}{1.3mm}{
  \begin{tabular}{lcccccc}
    \toprule
    \multirow{4}{*}{Models} & 
    \multicolumn{2}{c}{\multirow{2}{*}{\textbf{VoxCeleb1-O}}} & 
    \multicolumn{2}{c}{\multirow{2}{*}{\textbf{VoxCeleb1-E}}} &  \multicolumn{2}{c}{\multirow{2}{*}{\textbf{VoxCeleb1-H}}} \\ \\
    \cline{2-7} 
    & \multirow{2}{*}[-2pt]{\textbf{EER}} &  \multirow{2}{*}[-2pt]{$\textbf{DCF}_\textbf{0.01}$} & \multirow{2}{*}[-2pt]{\textbf{EER}} & \multirow{2}{*}[-2pt]{$\textbf{DCF}_\textbf{0.01}$} & 
     \multirow{2}{*}[-2pt]{\textbf{EER}} & \multirow{2}{*}[-2pt]{$\textbf{DCF}_\textbf{0.01}$} \\ \\
    \midrule
    A0      & 1.4310 & 0.1219   & 1.2900 & 0.1207   & 2.1700 & 0.1864 \\
    Var \textit{a}  & 1.3940 & 0.1185   & 1.3100 & 0.1256   & 2.2100 & 0.1913 \\
    Var \textit{b}    & 1.3890 & 0.1140   & 1.2980 & 0.1236   & 2.1802 & 0.1889\\
    Var \textit{c} & 1.3091 & 0.1169 & 1.3110 & 0.1265 & 2.2213 & 0.1937 \\
    Var \textit{d} & \textbf{1.2671} & \textbf{0.1039} & \textbf{1.2602} & \textbf{0.1181} & \textbf{2.1126} & \textbf{0.1850} \\
    Var \textit{e} & 1.4263 & 0.1320 & 1.3164 & 0.1266 & 2.2201 & 0.1940\\
    Var \textit{f} & \textbf{1.0821} & \textbf{0.1006} & \textbf{1.1204} & \textbf{0.1067} & \textbf{1.9342} & \textbf{0.1665}\\
    \bottomrule
  \end{tabular}}
  \label{tab:a0}
\end{table}
As we mentioned in Section \ref{variants}, the RepVGG block is task-specific. To compare our proposed variants with the original RepVGG structure, we selected RepVGG-A0 as our base model since it has a much faster training speed when compared with RepVGG-A1 and RepVGG-A2. To find out the most suitable re-parameterizable structure in speaker verification, we trained all the variants mentioned above. All the performances were presented in \textbf{Table \ref{tab:a0}}. As for Var \textit{a}, it is rather intriguing that replacing the original $1\times 1$ CONV-BN with an extra $3\times 3$ CONV-BN bring performance decay on some complex test sets like VoxCeleb1-E and VoxCeleb1-H even the training time state has more parameters (larger branch capacity). We believed that this extra $3\times 3$ CONV-BN structure tended to learn a representation that was similar to the main $3\times3$ CONV-BN. The lack of feature diversity caused the performance decay. Furthermore, Var \textit{d} performed the best among all models (Var \textit{f} not included) on all test sets. This structure added a $3\times 3$ CONV-BN after the original $1\times 1$ CONV-BN. On the contrary, Var \textit{e} which consists of a $1\times1$ CONV-BN and an average pooling performed the worst. Both these two structures had operators that control the balance between the branch diversity and branch capacity.
\begin{figure}[h]
  \centering
  \includegraphics[width=0.4\textwidth,height=0.2\textwidth]{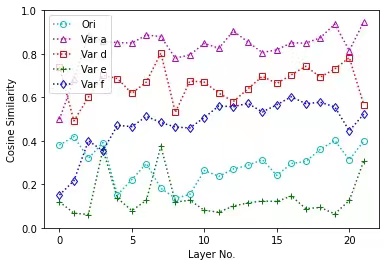}
  \caption{Branch similarity of each model. The layer number starts from the initial block to the final one.}
  \label{fig:similarity}
\end{figure}
\begin{table*}[t]
  \centering
  \caption{Performances of the RepSPKNet. RSBA denotes RepSPK-A Block and RSBB denotes RepSPK-B block. - $3 \times 3$ + $5 \times 5$ means replacing the main $3\times 3$ CONV-BN branch with a $5\times 5$ CONV-BN. The ECAPA denotes the ECAPA-TDNN(C=2048), and its results are referred from \cite{thienpondt2020idlab}. The ResNet34 model denotes our implementation of SOTA ResNet system. No score normalization is adopted except that the ECAPA system reports performances using adaptive s-norm.}
  \setlength{\tabcolsep}{7.6mm}{
  \begin{tabular}{lcccccc}
    \toprule
    \multirow{4}{*}{Models} & 
    \multicolumn{2}{c}{\multirow{2}{*}{\textbf{VoxCeleb1-O}}} & 
    \multicolumn{2}{c}{\multirow{2}{*}{\textbf{VoxCeleb1-E}}} &  \multicolumn{2}{c}{\multirow{2}{*}{\textbf{VoxCeleb1-H}}} \\ \\
    \cline{2-7} 
    & \multirow{2}{*}[-2pt]{\textbf{EER}} &  \multirow{2}{*}[-2pt]{$\textbf{DCF}_\textbf{0.01}$} & \multirow{2}{*}[-2pt]{\textbf{EER}} & \multirow{2}{*}[-2pt]{$\textbf{DCF}_\textbf{0.01}$} & 
     \multirow{2}{*}[-2pt]{\textbf{EER}} & \multirow{2}{*}[-2pt]{$\textbf{DCF}_\textbf{0.01}$} \\ \\
    \midrule
    ECAPA & 0.8600 & 0.0960 & 1.0800 & 0.1223 & 2.0100 & 0.2004\\
    ResNet34 & 1.0498 & 0.1045  & 1.0587 & 0.1008 & 1.8456 & 0.1619 \\
    \midrule
    A0      & 1.4310 & 0.1219   & 1.2900 & 0.1207   & 2.1700 & 0.1864 \\
    RSBA-A0 & \textbf{1.2671} & \textbf{0.1039} & \textbf{1.2602} & \textbf{0.1181} & \textbf{2.1126} & \textbf{0.1850} \\
    RSBB-A0 & \textbf{1.0821} & 0.1006 & 1.1204 & \textbf{0.1067} & 1.9342 & \textbf{0.1665}\\
    \ \ \ - $3 \times 3$ + $5 \times 5$  & 1.1771 & \textbf{0.0982} & \textbf{1.1041} & 0.1081  & \textbf{1.8960} & 0.1688 \\
    \midrule
    A1      & 1.2141 & 0.0913   & 1.1593 & 0.1054   & 1.9347 & 0.1655 \\
    RSBA-A1 & \textbf{1.1503} & \textbf{0.0872} & \textbf{1.1295} & \textbf{0.1013} & \textbf{1.8902} & \textbf{0.1605} \\
    RSBB-A1 & \textbf{0.9650} & \textbf{0.0795} & \textbf{1.0359} & \textbf{0.0936} & \textbf{1.7602} & \textbf{0.1512}\\
    \midrule
    A2      & 0.9546 & 0.0831   & 1.0143 & 0.0926   & 1.7149 & 0.1465 \\
    RSBA-A2 & \textbf{0.9122} & \textbf{0.0801} & \textbf{0.9939} & \textbf{0.0916} & \textbf{1.6809} & \textbf{0.1431} \\
    RSBB-A2 & \textbf{0.8430} & \textbf{0.0775} & \textbf{0.9637} & \textbf{0.0907} & \textbf{1.5982} & \textbf{0.1374}\\
    \bottomrule
  \end{tabular}}
  \label{tab:repspk}
\end{table*}

To verify that the multi-branch structure's performance depends on the trade-off between branch diversity and branch capacity, we designed a branch Var \textit{f} as shown in \textbf{Fig. \ref{fig:variants}}. This structure is a $3\times3$ CONV-BN with dilation as 2. This dilated convolution ensures diversity by constraining a different input and receptive field from the custom convolution of the main branch, and it also ensures capacity by increasing the parameters. We used the cosine similarity between the outputs of the main branch and our proposed branch to represent the branch similarity. The branch similarity of each layer was presented in \textbf{Fig. \ref{fig:similarity}}. Var \textit{a}, obviously had the highest branch similarities (around 0.9) as we speculated. Var \textit{e}, on the contrary, had the lowest similarities (around 0.2). Both the two structures showed performance decay. 
The other two variants had similarities around \textbf{0.5} and outperformed the base model. Moreover, Var \textit{f} with similarities closer to 0.5 achieved a relative 10.9\% EER and a relative 10.5\% minDCF improvement compared to the base RepVGG-A0 model.

\subsubsection{RepSPKNet architecture}
\begin{figure}[h]
  \centering
  \includegraphics[width=0.482\textwidth,height=0.21\textwidth]{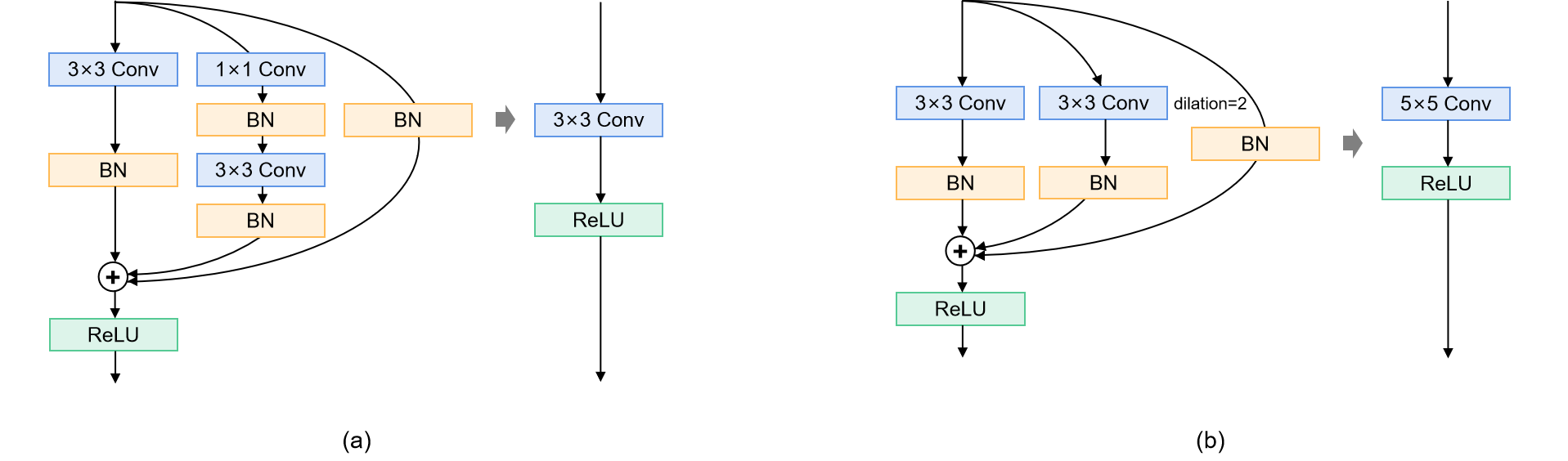}
  \caption{RepSPK block (RSB). (a) denotes the RSBA block and (b) denotes RSBB block. The ID-BN exists when input channel equals output channel.}
  \label{fig:RSB}
\end{figure}
It is easy to prove that a $3\times 3$ convolution with a dilation of 2 can be transformed to a $5\times 5$ convolution. According to the transformations provided by Section.\ref{sec:rep}, this Var \textit{f} structure can also be re-parameterized to a single $5\times5$ custom convolution. Based on the results, we proposed the final architecture of the RepSPKNet. As depicted in \textbf{Fig. \ref{fig:RSB}}, two blocks were presented. The RSBA block was composed of a $3\times3$ CONV-BN, a $1\times 1$ CONV-BN followed by a $3\times 3$ CONV-BN, and an ID-BN. The RSBB block was composed of a $3\times3$ CONV-BN, a $3\times3$ CONV-BN with a dilation of 2, and an ID-BN. To verify the stability and transferability of our proposed architecture, we compared our RepSPKNet with the original RepVGG-A1 and RepVGG-A2. Here we only replaced the RepVGG block with the RepSPK block (RSB). The model consisted of RSBA blocks was called RepSPKNet-A and the other consisted of RSBB blocks was called RepSPKNet-B. We also conducted an ablation study of the RSBB structure by replacing the $3\times 3$ CONV-BN main branch with a $5\times 5$ CONV-BN. The results were presented in \textbf{Table \ref{tab:repspk}}. The RepSPKNet-A and RepSPKNet-B both outperformed their corresponding base model. Compared to the performance of RepVGG-A2 on VoxCeleb1-H, the RSBA-A2 achieved relative improvements of 2.0\% in EER and 2.3\% in minDCF. Moreover, the RSBB-A2 achieved relative improvements of 6.8\% in EER and 6.2\% in minDCF. The result demonstrated that our RepSPKNets can achieve SOTA performance in speaker verification.

\section{Conclusion}
\label{sec:conclusion}

In this paper, we proposed two blocks as \textbf{Fig.\ref{fig:RSB}} demonstrates. The RepSPKNet-A is composed of the RSBA block while the RepSPKNet-B is composed of the RSBB block. With the structral re-parameterization method, the RepSPKNet-A can be transformed to a stack of $3\times3$ convolution and ReLU and the RepSPKNet-B can be transformed to a stack of $5\times 5$ convolution and ReLU. Ablation studies on various variants indicated that the performance of multi-branch structure depended on the branch diversity and branch capacity, which was a heuristic principle for designing multi-branch models. Our proposed RepSPKNet outperformed the original RepVGG and achieved SOTA performance in speaker verification.

\vfill\pagebreak

\bibliographystyle{IEEEbib}
\bibliography{refs}

\end{document}